\documentclass[12pt]{article}
\usepackage{amssymb,amsmath,epsfig}

\begin{document}

\title{\bf Anisotropic Universe Models with Perfect Fluid and Scalar Field in $f(R,T)$ Gravity}

\author{Muhammad SHARIF \thanks{msharif.math@pu.edu.pk} and Muhammad ZUBAIR
\thanks{mzubairkk@gmail.com}\\
Department of Mathematics, University of the Punjab,\\
Quaid-e-Azam Campus, Lahore-54590, Pakistan.}

\date{}

\maketitle

\begin{abstract}
In this paper, we study the behavior of perfect fluid and massless
scalar field for homogeneous and anisotropic Bianchi type I universe
model in $f(R,T)$ gravity, where $R$ is the Ricci scalar and $T$ is
the trace of the energy-momentum tensor. We assume the variation law
of mean Hubble parameter to obtain exact solutions of the modified
field equations. The physical and kinematical quantities are
discussed for both models in future evolution of the universe. We
check the validity of null energy condition and conclude that our
perfect fluid solution can behave like phantom model. Finally, we
find that perfect fluid solutions correspond to massless scalar
field models.
\end{abstract}
{\bf Keywords:} $f(R,T)$ gravity; Dark energy; Exact solutions;
Massless scalar field.\\
{\bf PACS:} 04.20.Jb; 04.50.Kd; 95.36.+x.

\section{Introduction}

Over the last decade, the most significant progress in astrophysics
and cosmology is the observational evidence that the present
universe is undergoing a phase of accelerated expansion.
Observations from supernova type Ia (SNeIa)$^{{1})}$, cosmic
microwave background (CMB) anisotropies $^{{2})}$, large scale
structure$^{{3})}$, baryon acoustic oscillations$^{{4})}$ and weak
lensing$^{{5})}$ show that most of the cosmic energy density is
dominated by exotic energy source known as \emph{dark energy} (DE).
The DE is said to be responsible for this cosmic acceleration but
its nature is still an important challenge. There have been proposed
many candidates to address this issue. The cosmological constant or
vacuum energy is the simplest candidate which appears to fit the
observational data. However, despite of its success the $\Lambda$CDM
($\Lambda$-cold dark matter) model faces serious fine-tuning and
coincidence problems$^{{6})}$.

Currently there are two different approaches to address the cosmic
acceleration issue. One approach is to introduce various scalar
fields of matter in Einstein gravity such as quintessence, phantom
fields, tachyon field, Chaplygin gas$^{{7})}$ etc and also cosmic
fluids with anisotropic equation of state (EoS)$^{{8})}$. The other
approach is based on modification of the Einstein-Hilbert action to
get alternative theories of gravity such as $f(R)$ gravity$^{{9})}$,
$f(T)$ gravity$^{{10})}$, Gauss-Bonnet gravity$^{{11})}$. Harko et
al.$^{{12})}$ presented a new modification of Einstein Lagrangian by
introducing an arbitrary function of scalar curvature $R$ and trace
of the energy-momentum tensor $T$. The dependence of $T$ may be
introduced by exotic imperfect fluids or quantum effects (conformal
anomaly).

The action of $f(R,T)$ theory of gravity is given by$^{{12})}$
\begin{equation}\label{1}
\mathcal{A}=\frac{1}{16{\pi}}\int{f(R,T)\sqrt{-g}dx^4}+\int\mathcal{L}_{m}\sqrt{-g}dx^4,
\end{equation}
where $G=c=1$, $g$ is the determinant of the metric tensor
$g_{\mu\nu}$ and $\mathcal{L}_{m}$ determines matter contents of the
universe. The energy-momentum tensor of matter is defined
as$^{{13})}$
\begin{equation}\label{2}
T_{{\mu}{\nu}}=-\frac{2}{\sqrt{-g}}\frac{\delta(\sqrt{-g}
{\mathcal{\mathcal{L}}_{m}})}{\delta{g^{{\mu}{\nu}}}}.
\end{equation}
In fact, this modified gravity is the generalization of $f(R)$
gravity and is based on coupling between matter and geometry. The
corresponding field equations have been derived in metric formalism
for several particular cases of $f(R,T)$ gravity$^{{12})}$. They
have also explored possibility of reconstructing the FRW cosmologies
by an appropriate choice of $f(T)$ for the model $f(R,T)=R+2f(T)$.
This model has been used to investigate perfect fluid solutions of
spatially homogeneous and anisotropic Bianchi models$^{{14})}$. In
recent papers$^{{15})}$, the cosmological reconstruction of $f(R,T)$
gravity models have been investigated. We have also explored the
validity of first and second laws of thermodynamics in $f(R,T)$
gravity$^{{16})}$.

The existence of anomalies in CMB still requires an intense debate
which stimulated our interest in anisotropic Bianchi models. Many
authors$^{{17-19})}$ have studied spatially homogeneous and
anisotropic Bianchi models in the context of modified theories of
gravity. Sharif and Shamir$^{{17})}$ explored vacuum and non-vacuum
solutions of Bianchi I and V universe models in $f(R)$ gravity.
Bianchi models being anisotropic are useful to study isotropic
behavior of the universe with the passage of time. Sharif and Kausar
$^{{18})}$ investigated the isotropic behavior of Bianchi III model
in $f(R)$ gravity. Scalar field can play a vital role to explain the
cosmic acceleration which has widely been studied in $f(R)$ gravity
$^{{20})}$. As $f(R,T)$ involves coupling between matter and
geometry, so considering scalar field as a source may provide some
new insights.

In this work, we study perfect fluid and massless scalar field
solutions of locally rotationally symmetric (LRS) Bianchi type I
(BI) universe in $f(R,T)$ gravity. Our aim is to find exact
solutions of the field equations in this theory and discuss the
future evolution of the universe. The paper is organized as follows:
In section \textbf{2}, we formulate the field equations for LRS BI
universe in $f(R,T)$ gravity. Section \textbf{3} provides solutions
of the field equations and investigates physical behavior of models
and kinematical parameters. Section \textbf{4} contains solutions of
the field equations for massless scalar field. Finally, section
\textbf{5} is devoted to discussion and conclusion of the results.

\section{$f(R,T)$ Gravity and Bianchi I Universe}

Variation of the action (\ref{1}) with respect to the metric tensor
results in the following field equations of $f(R,T)$ gravity
\begin{eqnarray}\label{3}
&&R_{{\mu}{\nu}}f_{R}(R,T)-\frac{1}{2}g_{{\mu}{\nu}}f(R,T)+(g_{{\mu}{\nu}}
{\Box}-{\nabla}_{\mu}{\nabla}_{\nu})f_{R}(R,T)\nonumber\\&=&8{\pi}
T_{{\mu}{\nu}}-f_{T}(R,T)T_{{\mu}{\nu}}-f_{T}(R,T)\Theta_{{\mu}{\nu}},
\end{eqnarray}
where $f_{R}(R,T)$ and $f_{T}(R,T)$ denote derivatives of $f(R,T)$
with respect to $R$ and $T$ respectively;
${\Box}={\nabla}_{\mu}{\nabla}^{\mu}$, ${\nabla}_{\mu}$ is the
covariant derivative associated with the Levi-Civita connection of
the metric tensor and $\Theta_{{\mu}{\nu}}$ is defined by
\begin{equation}\label{4}
\Theta_{{\mu}{\nu}}=\frac{g^{\alpha{\beta}}{\delta}T_{{\alpha}{\beta}}}
{{\delta}g^{\mu{\nu}}}=-2T_{{\mu}{\nu}}+g_{\mu\nu}\mathcal{L}_m
-2g^{\alpha\beta}\frac{\partial^2\mathcal{L}_m}{{\partial}
g^{\mu\nu}{\partial}g^{\alpha\beta}}.
\end{equation}
Since the field equations depend on $\Theta_{{\mu}{\nu}}$, so each
form of $\mathcal{L}_m$ would generate a specific set of the field
equations. The choice of $f(R,T){\equiv}f(R)$ results in the field
equations of $f(R)$ gravity. The trace of equation (\ref{3}) is
given by
\begin{equation}\label{5}
Rf_{R}(R,T)+3{\Box}f_{R}(R,T)-2f(R,T)=8{\pi}
T-f_{T}(R,T)T-f_{T}(R,T)\Theta,
\end{equation}
where $\Theta=\Theta_{\mu}^{\mu}$. This equation leads to $f(R,T)$
in terms of its derivatives and matter contents of the universe as
\begin{eqnarray}\label{5a}
f(R,T)=\frac{1}{2}\left[Rf_R(R,T)+3{\Box}f_{R}(R,T)-8{\pi}
T+f_{T}(R,T)T+f_{T}(R,T)\Theta\right].
\end{eqnarray}
Eliminating $f(R,T)$ from Eqs.(\ref{3}) and (\ref{5}), we obtain
\begin{eqnarray}\label{6}
&&(R_{{\mu}{\nu}}-\frac{1}{4}Rg_{\mu\nu})f_{R}(R,T)+(\frac{1}{4}g_{{\mu}{\nu}}
{\Box}-{\nabla}_{\mu}{\nabla}_{\nu})f_{R}(R,T)=8{\pi}
(T_{{\mu}{\nu}}\nonumber\\&-&\frac{1}{4}Tg_{\mu\nu})-f_{T}(R,T)(T_{{\mu}{\nu}}-\frac{1}{4}Tg_{\mu\nu})
-f_{T}(R,T)(\Theta_{{\mu}{\nu}}-\frac{1}{4}{\Theta}g_{\mu\nu}).
\end{eqnarray}

The line element for the spatially homogeneous and anisotropic LRS
BI spacetime is given by
\begin{equation}\label{7}
ds^{2}=dt^2-A^2(t)dx^2-B^2(t)(dy^2+dz^2),
\end{equation}
where the scale factors $A$ and $B$ are functions of cosmic time $t$
only. For $A(t)=B(t)=a(t)$, this reduces to the flat FRW spacetime.
The energy-momentum tensor of perfect fluid is
\begin{equation*}
T_{{\mu}{\nu}}=({\rho}+p)u_{\mu}u_{\nu}-pg_{{\mu}{\nu}},
\end{equation*}
where $u_{\mu}$ is the four velocity of the fluid, $\rho$ and $p$
denote the energy density and pressure, respectively. We assume
equation of state (EoS) $p=\omega\rho$ with $\omega$ being a
constant. If $\omega=-1$, we have $\Lambda$CDM model, $\omega>-1$
represents quintessence and $\omega<-1$ denotes phantom era. The
matter lagrangian can be assumed as $\mathcal{L}_{m}=-p$. Using
Eq.(\ref{4}), $\Theta_{{\mu}{\nu}}$ becomes
\begin{equation}\label{9}
\Theta_{{\mu}{\nu}}=-2T_{{\mu}{\nu}}-pg_{{\mu}{\nu}}.
\end{equation}
Consequently, the field equations (\ref{6}) lead to
\begin{eqnarray}\label{10}
&&(R_{{\mu}{\nu}}-\frac{1}{4}Rg_{\mu\nu})f_{R}(R,T)+(\frac{1}{4}g_{{\mu}{\nu}}
{\Box}-{\nabla}_{\mu}{\nabla}_{\nu})f_{R}(R,T)=8{\pi}
(T_{{\mu}{\nu}}\nonumber\\&-&\frac{1}{4}Tg_{\mu\nu})+f_{T}(R,T)
(T_{{\mu}{\nu}}+pg_{\mu\nu}) -\frac{\lambda}{4}(\rho+p)g_{\mu\nu}.
\end{eqnarray}

We consider the function $f(R,T)$ of the form$^{{15})}$
\begin{equation}\label{11}
f(R,T)=f(R)+\lambda{T},
\end{equation}
where $\lambda$ is a constant, $f(R)$ is an arbitrary function of
$R$ and $T$, the trace of energy-momentum tensor is a correction to
$f(R)$ theory. This choice involves explicit coupling of matter and
geometry, so it can produce significant results.  The field
equations (\ref{10}) for BI universe model take the form
\begin{eqnarray}\label{12}
&&\left(\frac{\ddot{A}}{A}+2\frac{\ddot{B}}{B}-2\frac{\dot{A}\dot{B}}{AB}
-2\frac{\dot{B}^2}{B^2}\right)F+\frac{3}{2}\ddot{F}-\frac{1}{2}\left(\frac{\dot{A}}{A}
+2\frac{\dot{B}}{B}\right)\dot{F}\nonumber\\&=&-\frac{3}{2}(8\pi+\lambda)(\rho+p),\\\label{13}
&&\left(\frac{\ddot{A}}{A}-2\frac{\ddot{B}}{B}+2\frac{\dot{A}\dot{B}}{AB}
-2\frac{\dot{B}^2}{B^2}\right)F-\frac{1}{2}\ddot{F}+\left(\frac{3}{2}\frac{\dot{A}}{A}
-\frac{\dot{B}}{B}\right)\dot{F}\nonumber\\&=&\frac{1}{2}(8\pi+\lambda)(\rho+p),\\\label{14}
&&\left(\frac{\dot{B}^2}{B^2}-\frac{\ddot{A}}{A}\right)F-\frac{1}{2}\ddot{F}-
\frac{1}{2}\left(\frac{\dot{A}}{A}-2\frac{\dot{B}}{B}\right)\dot{F}
=\frac{1}{2}(8\pi+\lambda)(\rho+p),
\end{eqnarray}
where $F(R)$ denotes derivative of $f(R)$ with respect to the Ricci
scalar $R$ and
\begin{equation}\label{15}
R=-2\left(\frac{\ddot{A}}{A}+2\frac{\ddot{B}}{B}+2\frac{\dot{A}\dot{B}}{AB}
+\frac{\dot{B}^2}{B^2}\right).
\end{equation}

Now we define some physical quantities for BI model which are
important in cosmological observations. The average scale factor,
volume, expansion and shear scalars are defined as
\begin{equation}\label{16}
V=a^3=AB^2, \quad
\theta=u_{;a}^a=\frac{\dot{A}}{A}+2\frac{\dot{B}}{B},\quad
\sigma^2=\frac{1}{2}\sigma_{ab}\sigma^{ab}=
\frac{1}{3}\left[\frac{\dot{A}}{A}-\frac{\dot{B}}{B}\right]^2.
\end{equation}
The anisotropy parameter of the expansion is characterized by the
mean and directional Hubble parameters defined as
\begin{equation}\label{19}
\Delta=\frac{2}{9}\left(\frac{H_x-H_y}{H}\right)^2,
\end{equation}
where
\begin{equation*}
H=({\ln}{a}\dot)=\frac{1}{3}\left(\frac{\dot{A}}{A}+2\frac{\dot{B}}{B}\right)
\end{equation*}
is the mean Hubble parameter and $H_i(i=1,2,3)$ represent the
directional Hubble parameters on $x,~y$ and $z$ axes respectively,
and are given by
\begin{equation*}
H_x=\frac{\dot{A}}{A}, \quad H_y=H_z=\frac{\dot{B}}{B}.
\end{equation*}
The anisotropy of the expansion results in isotropic expansion of
the universe in the limit of $\Delta\longrightarrow0$. The
deceleration parameter is defined as
\begin{equation}\label{20}
q=\frac{d}{dt}\left(\frac{1}{H}\right)-1,
\end{equation}
which can be used to explain the transition from past deceleration
to the present accelerating epoch$^{{21})}$.

\section{Solution of the Field Equations}

To solve the field equation in $f(R,T)$ gravity, we assume the
variation law of mean Hubble parameter defined by the relation
\begin{equation}\label{21}
H=la^{-m}=l(AB^2)^{-m/3}, \quad l>0, \quad m\geqslant0.
\end{equation}
Berman$^{{22})}$ proposed this law for spatially homogeneous and
isotropic FRW spacetime which yields constant value of the
deceleration parameter. In recent papers$^{{16-19})}$, a similar law
is used to generate exact solutions for the homogeneous and
anisotropic Bianchi models in modified theories of gravity. Using
$H$ and $V$ for the BI model in Eq.(\ref{21}), we obtain two
different volumetric expansion laws
\begin{eqnarray}\label{22}
V&=&c_1e^{3lt},\quad m=0,\\\label{23}
V&=&(mlt+c_2)^{3/m},\quad
m\neq0,
\end{eqnarray}
where $c_1$ and $c_2$ are positive constants. Equation (\ref{22})
corresponds to de Sitter expansion with the scale factor being
increasing function of cosmic time as $a(t)=a_0e^{Ht}$, $H=l=$
constant. The second volumetric expansion law represents power law
model with scale factor $a(t)=a_0t^n$. If $0<n<1$, then power law
solution is accelerating and for $n>1$, it exhibits decelerating
behavior. Subtracting Eq.(\ref{14}) from (\ref{13}) with some
manipulation, it follows that
\begin{equation}\label{24}
H_x-H_y=\frac{k}{VF},
\end{equation}
where $k$ is a positive constant. Using Eqs.(\ref{16}) and
(\ref{24}) in (\ref{19}), it turns out
\begin{equation}\label{25}
\Delta=6\left(\frac{\sigma}{\theta}\right)^2=\left(\frac{k}{\sqrt{3}\dot{V}F}\right)^2.
\end{equation}
In the following, we discuss above two cases separately.

\subsection{Perfect Fluid Model When $m=0$}

Here, the spatial volume of the universe for exponential expansion
is given by Eq.(\ref{22}). Using this value of $V$ in Eq.(\ref{24}),
we can write the scale factors as
\begin{equation}\label{27}
A=c_1^{1/3}c_3^{2/3}e^{lt+\frac{2k}{3}\int\frac{1}{VF}dt}, \quad
B=c_1^{1/3}c_3^{-1/3}e^{lt-\frac{k}{3}\int\frac{1}{VF}dt}.
\end{equation}
To find the explicit solution of the field equations, we assume a
relation between $F$ and $a$ as $F\propto{a^{n}}$$^{{17-18})}$,
which implies that
\begin{equation*}
F={\alpha}e^{nlt},
\end{equation*}
where $\alpha$ is the proportionality constant and $n$ is any
arbitrary constant. As we are interested to discuss the exponential
and power law expansions, so it would be useful to assume unknown
$F$ in terms of these expansion laws. This assists to reconstruct
the $f(R,T)$ gravity depending on the choice of the scale factor.
Using this value of $F$ in Eq.(\ref{27}), we obtain
\begin{equation}\label{30}
A=c_1^{1/3}c_3^{2/3}e^{lt-\frac{2k}{3{\alpha}l(n+3)}e^{-(n+3)lt}},
\quad
B=c_1^{1/3}c_3^{-1/3}e^{lt+\frac{k}{3{\alpha}l(n+3)}e^{-(n+3)lt}}.
\end{equation}

For $n>-3$, we observe that the scale factors $A(t)$ and $B(t)$ have
constant values at initial epoch which imply that the model has no
initial singularity, while these diverge in future evolution of the
universe. When $n<-3$, the scale factors increase with time and
approach to very large values as $t\rightarrow{\infty}$. For $n=-3$,
the model represents similar behavior in every direction. The
directional, mean Hubble parameters and anisotropy parameter of
expansion turn out to be
\begin{eqnarray*}
&&H_x=l+\frac{2k}{3\alpha}e^{-(n+3)lt}, \quad
H_y=H_z=l-\frac{k}{3\alpha}e^{-(n+3)lt},\quad H=l,\\\nonumber
&&{\Delta}=\frac{2k^2}{9l^2{\alpha}^2}e^{-2(n+3)lt}.
\end{eqnarray*}
\begin{figure}
\centering \epsfig{file=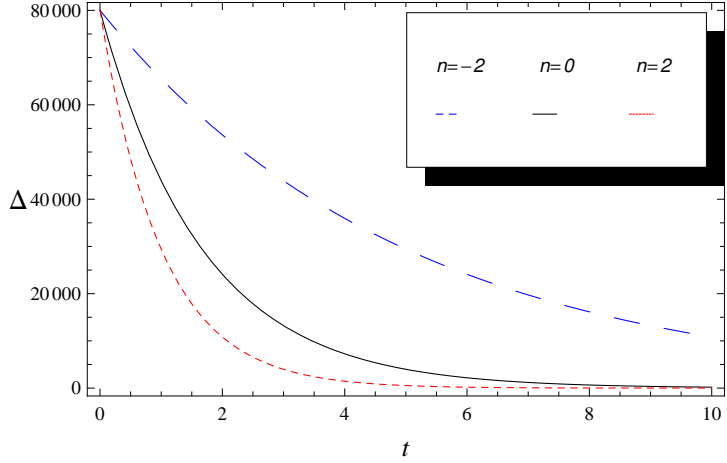} \caption{Plot of $\Delta$ versus
cosmic time for different values of $n$. We set $l=0.1$, $k=3$, and
$\alpha=0.05$. (Colour online)}
\end{figure}
The mean Hubble parameter is found to be constant whereas the
directional Hubble parameters are dynamical. For $n>-3$, $H_x$ and
$H_y$ become constant at $t=0$ as well as for
$t\rightarrow{\infty}$. These parameters vary from $H$ by some
constant at $t=0$ and coincide for later times of the universe. As
the constant being positive (negative), it would increase (decrease)
expansion on the $x$-axis and it decreases (increases) expansion on
$y$ and $z$ axes. For $n=-3$, $H_x$ will increase from $H$ by a
constant factor $\frac{2k}{3\alpha}$, while parameters $H_y$, $H_z$
will decrease by a factor $\frac{k}{3\alpha}$. The anisotropy
parameter of expansion is found to be finite for earlier times of
the universe and vanishes as $t\rightarrow{\infty}$ for $n>-3$. The
plot of $\triangle$ versus cosmic time $t$ is shown in Figure
\textbf{1}.

The deceleration parameter, expansion and shear scalars are given by
\begin{eqnarray}\label{34}
q=-1, \quad \theta=3l=3H, \quad
\sigma^2=\frac{k^2}{3\alpha^2}e^{-2(n+3)lt}.
\end{eqnarray}
The volume $V$ of the universe is an exponential function which
expands with the increase in time and becomes infinitely large for
later times of the universe. Also, the expansion scalar is constant
for all times and hence the model would favor the uniform expansion.
The deceleration parameter $(q=-1)$ allows the existence of
accelerating model for this case which is in agreement with the
current observations of SNeIa and CMB$^{{1-2})}$.

Using Eqs.(\ref{30}) in Eqs.(\ref{12})-(\ref{14}), we obtain the
following relation of energy density and pressure
\begin{equation}\label{36}
\rho+p=\frac{-1}{3\alpha(8\pi+\lambda)}\left[2k^2e^{nlt-2(n+3)lt}+3b_1e^{nlt}\right],
\end{equation}
where $b_1=n(n-1)l^2\alpha^2$. This shows that the null energy
condition (NEC) is violated, \emph{i.e.}, $\rho+p<0$ which implies
that $\omega<-1$. Matter component with $\omega<-1$ is named as
"phantom energy" and is a possible candidate of the present
accelerated expansion. The phantom regime favors recent
observational cosmology of accelerated cosmic expansion. The
behavior of NEC for different values of $\alpha$ is displayed in
Figure \textbf{2}, which shows that NEC is violated for positive
values of $\alpha$. Thus, we assume $\alpha>0$ for phantom universe.
Equation (\ref{36}) implies the following dynamical variables of the
perfect fluid
\begin{figure}
\centering \epsfig{file=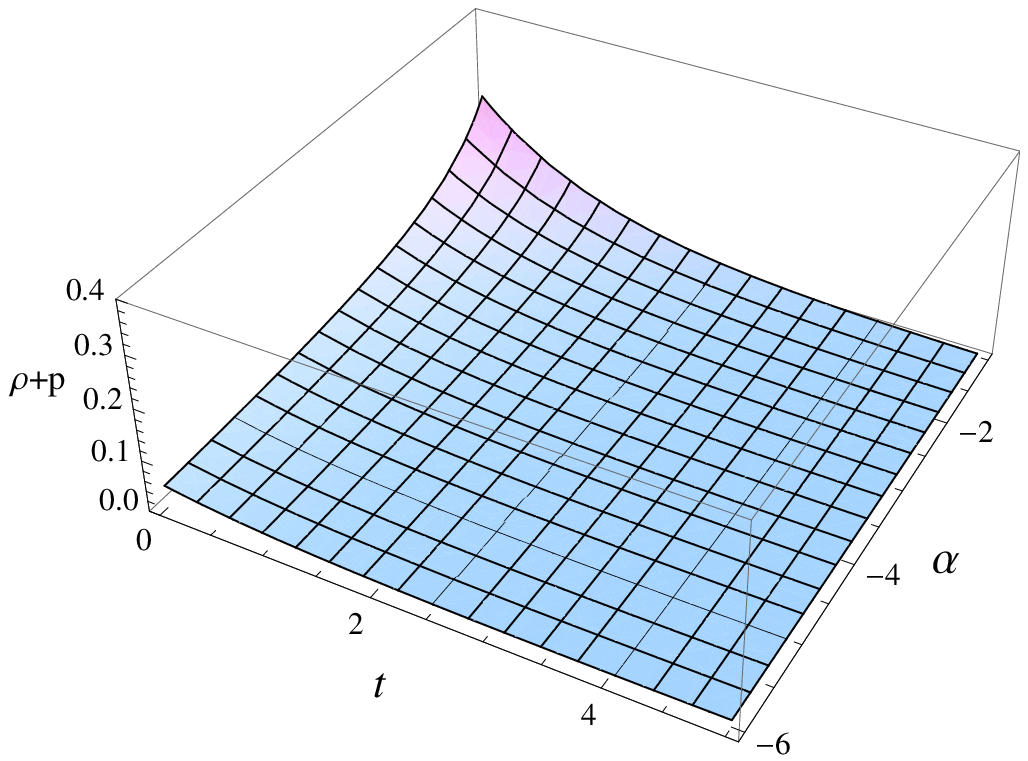,width=.50\linewidth,
height=2.2in}\epsfig{file=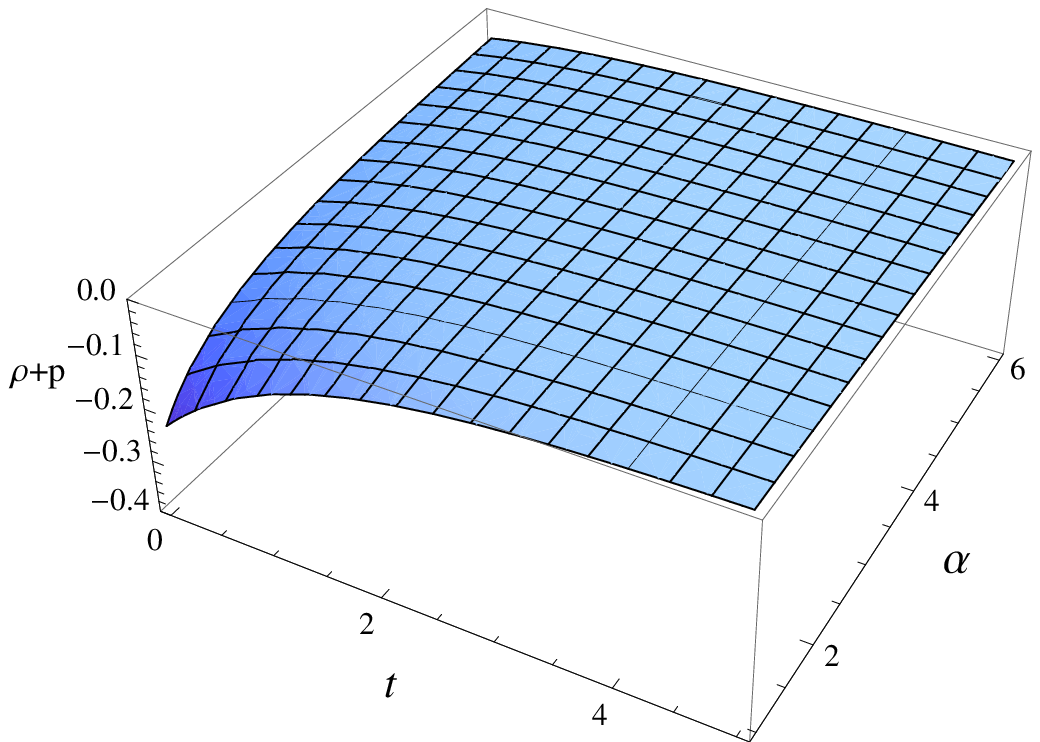,width=.50\linewidth,
height=2.2in} \caption{Evolution of NEC for $n=2$. The left graph
shows that NEC is satisfied for $\alpha<0$ and it is violated for
$\alpha>0$ at the right side. We set $l=\lambda=0.1$ and $k=3$.
(Colour online)}
\end{figure}
\begin{eqnarray}\label{37}
&&\rho=\frac{-1}{3\alpha(1+\omega)(8\pi+\lambda)}\left[2k^2e^{nlt-2(n+3)lt}
+3b_1e^{nlt}\right], \\\label{38}
&&p=\frac{-\omega}{3\alpha(1+\omega)(8\pi+\lambda)}\left[2k^2e^{nlt-2(n+3)lt}
+3b_1e^{nlt}\right].
\end{eqnarray}

For the phantom evolution of the universe, $\rho$ decreases with
cosmic time and approaches to zero as $t\rightarrow{\infty}$ in the
range of $-6<n\leq0$. When $n<-6$, $\rho$ increases as time goes
from zero to infinity and hence diverges. Figure \textbf{3} shows
that $\rho$ decreases for $n=-1, -4$ and becomes uniform for $n=-6$.
However, the value of $n=-7$ shows increasing $\rho$ for the future
evolution of the universe. If $n>0,~\rho$ decreases with time but
for large values of $n$, it shows bouncing behavior as shown in
Figure \textbf{3}. For this model, the scalar curvature $R$ and
$f(R,T)$ are given by
\begin{eqnarray*}
R&=&-\frac{2}{3\alpha^2}\left[18l^2\alpha^2+k^2e^{-2(n+3)lt}\right],\\\nonumber
f(R,T)&=&\frac{\alpha}{2}(R+3l(n^2l+3))e^{nlt}
+\frac{8\pi(1-3\omega)+\lambda(1-\omega)}{6\alpha(8\pi+\lambda)(1+\omega)}\\\nonumber
&\times&(2k^2e^{-(n+6)lt}+3b_1e^{nlt}).
\end{eqnarray*}
\begin{figure}
\centering \epsfig{file=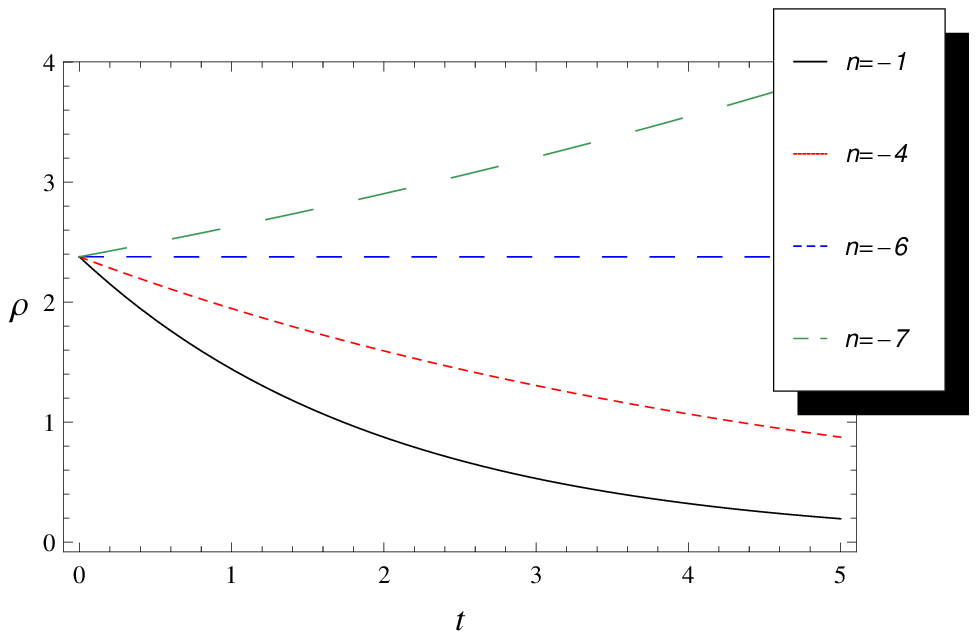,width=.50\linewidth,
height=2.2in}\epsfig{file=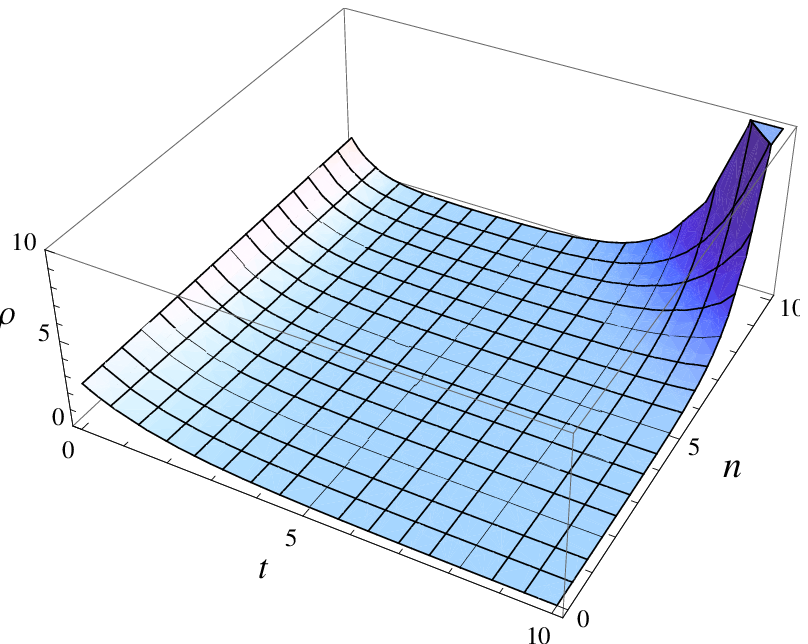,width=.50\linewidth,
height=2.2in} \caption{The left graph shows the behavior of $\rho$
for $-6<n\leq0$ and $n<-6$, while the right graph presents the
evolution of $\rho$ for $n>0$. We set $l=\lambda=0.1$, $k=3$ and
$\alpha=0.05$. (Colour online)}
\end{figure}

\subsection{Perfect Fluid Model When $m\neq0$}

For $m\neq0$, the spatial volume is given by Eq.(\ref{23}) and the
corresponding deceleration parameter is $q=m-1$. If $q>0$, the model
represents decelerating universe whereas $q<0$ indicates inflation.
To obtain the accelerated expansion model, we take $m<1$. Solving
the field equations (\ref{12})-(\ref{14}), the scale factors are
found to be
\begin{eqnarray}\nonumber
&&A=c_4^{2/3}(mlt+c_2)^{1/m}e^{\frac{2k}{3{\alpha}l(m-n-3)}(mlt+c_2)^{1-\frac{n+3}{m}}},\\\label{43}
&&B=c_4^{-1/3}(mlt+c_2)^{1/m}e^{\frac{k}{3{\alpha}l(n-m+3)}(mlt+c_2)^{1-\frac{n+3}{m}}}.
\end{eqnarray}
We discuss the evolution of the scale factors for two cases, i.e.,
$m>n+3$ and $m<n+3$ along with $0<m<1$. If $m>n+3$, the scale factor
$A$ increases with time whereas $B$ tends to zero. For $m<n+3$, the
behavior of scale factors is almost identical provided that $n$ is
always greater than $-3$ to keep $m$ positive. Substituting
Eq.(\ref{43}) in Eq.(\ref{19}), we get
\begin{eqnarray}\label{44}
&&H=l(mlt+c_2)^{-1}, \quad
H_x=l(mlt+c_2)^{-1}+\frac{2k}{3\alpha}(mlt+c_2)^{-\frac{n+3}{m}},\\\label{45}
&&H_y=H_z=l(mlt+c_2)^{-1}-\frac{k}{3\alpha}(mlt+c_2)^{-\frac{n+3}{m}},\\\label{46}
&&{\Delta}=\frac{2k^2}{9l^2{\alpha}^2}(mlt+c_2)^{-2\frac{(n+3)}{m}}.
\end{eqnarray}
\begin{figure}
\centering \epsfig{file=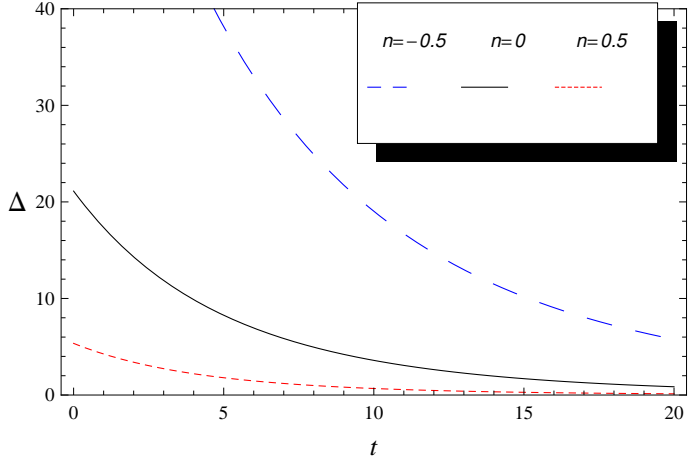} \caption{Plot of $\Delta$ versus
cosmic time $t$ for different values of $n$. We set $l=0.1$,
$k=c_2=3$, $m=0.9$ and $\alpha=0.05$. (Colour online)}
\end{figure}
The Hubble parameters $H,~H_x,~H_y$ and $H_z$ become constant at the
initial epoch. As $t\rightarrow{\infty}$, the values of these
parameters tend to zero for $n>-3$ and become infinite for $n<-3$.

If $n<-3,~\triangle$ increases with cosmic time whereas for $n>-3$,
its value decreases and may result to isotropic expansion in future
evolution of the universe (see Figure \textbf{4}). The expansion and
shear scalars turn out to be
\begin{equation}\label{48}
\theta=3l(mlt+c_2)^{-1},\quad
\sigma^2=\frac{k^2}{3\alpha^2}(mlt+c_2)^{-2\frac{(n+3)}{m}}.
\end{equation}
If we replace Eq.(\ref{43}) in Eqs.(\ref{12})-(\ref{14}), we obtain
\begin{eqnarray}\label{50}
&&\rho+p=\frac{-1}{3\alpha(8\pi+\lambda)}\left[2k^2(mlt+c_2)^{-\frac{(n+6)}{m}}
+3b_2(mlt+c_2)^{\frac{(n-2m)}{m}}\right],
\end{eqnarray}
where $b_2=(n(n-1)-m(n+2))l^2\alpha^2$. $\rho$ and $p$ are obtained
as follows
\begin{eqnarray}\label{51}
\rho=\frac{-1}{3\alpha(1+\omega)(8\pi+\lambda)}\left[2k^2(mlt+c_2)^{-\frac{(n+6)}{m}}
+3b_2(mlt+c_2)^{\frac{(n-2m)}{m}}\right], \\\label{52}
p=\frac{-\omega}{3\alpha(1+\omega)(8\pi+\lambda)}\left[2k^2(mlt+c_2)^{-\frac{(n+6)}{m}}
+3b_2(mlt+c_2)^{\frac{(n-2m)}{m}}\right].
\end{eqnarray}
\begin{figure}
\centering \epsfig{file=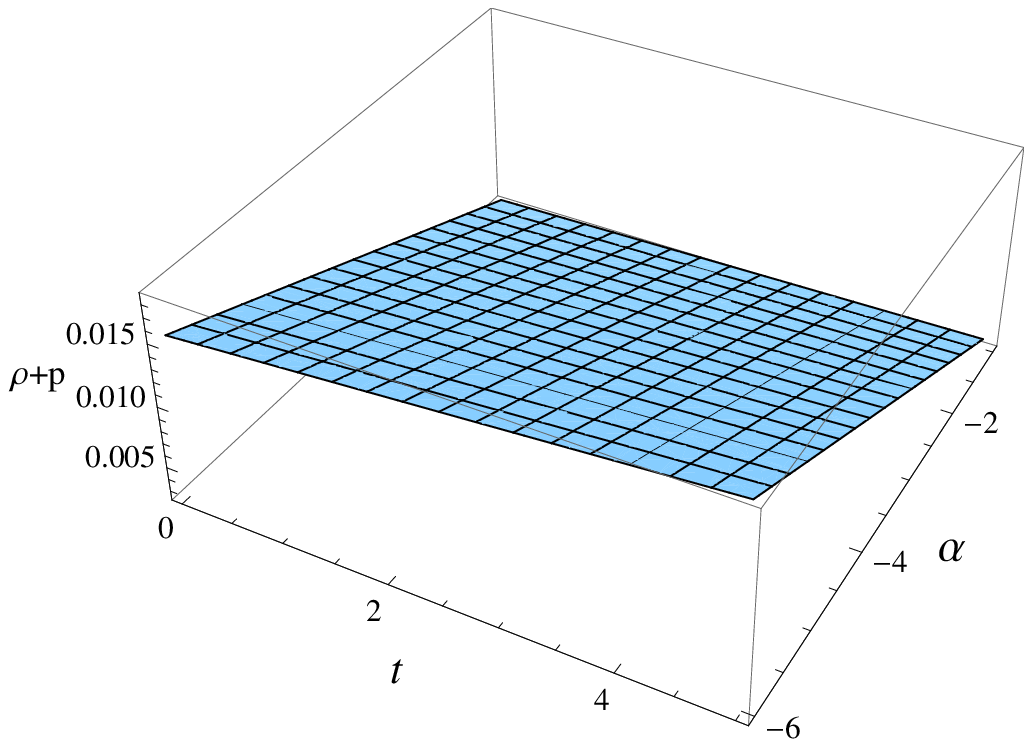,width=.50\linewidth,
height=2.2in}\epsfig{file=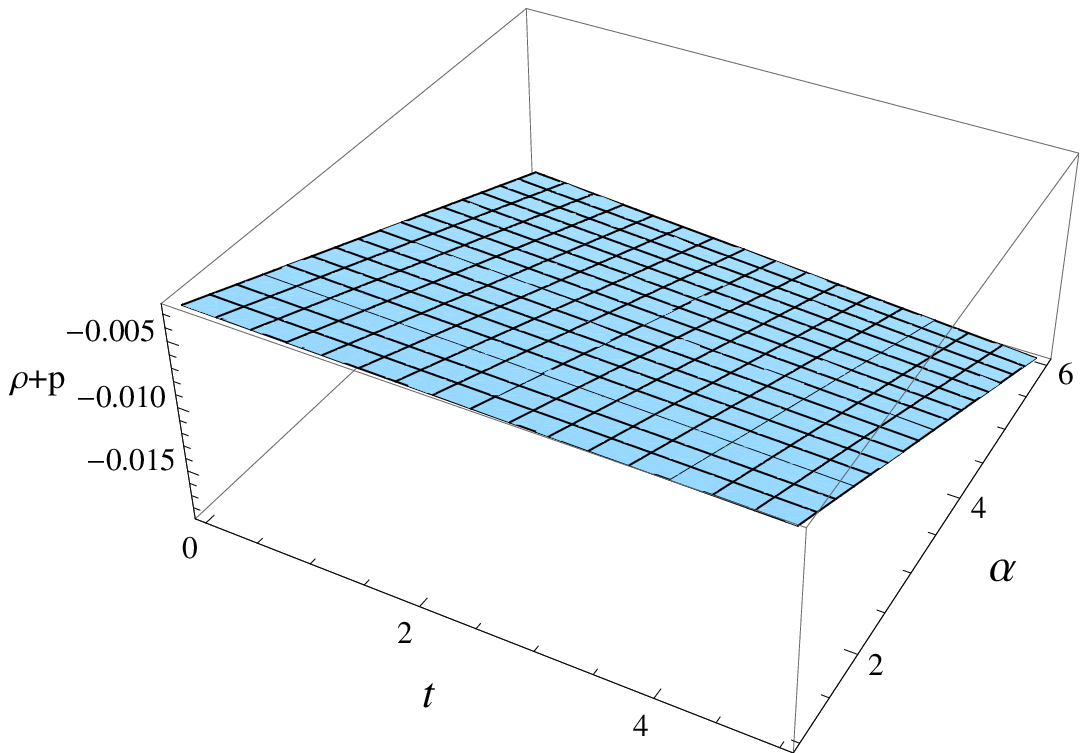,width=.50\linewidth,
height=2.2in} \caption{Behavior of NEC versus $\alpha$ for $n=3$.
The left part shows that NEC is satisfied for $\alpha<0$, while it
is violated for $\alpha>0$ shown on the right side.  We set
$l=\lambda=0.1$, $k=c_2=3$ and $m=0.9$. (Colour online)}
\end{figure}
\begin{figure}
\centering \epsfig{file=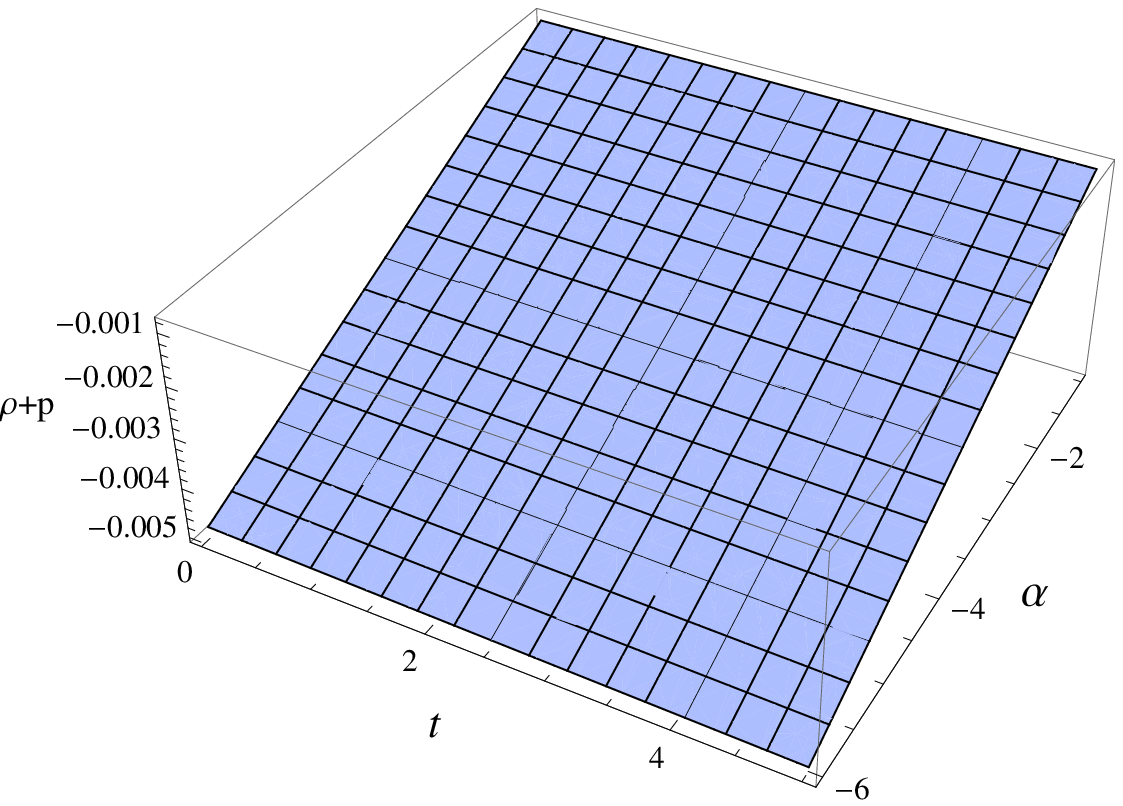,width=.50\linewidth,
height=2.2in}\epsfig{file=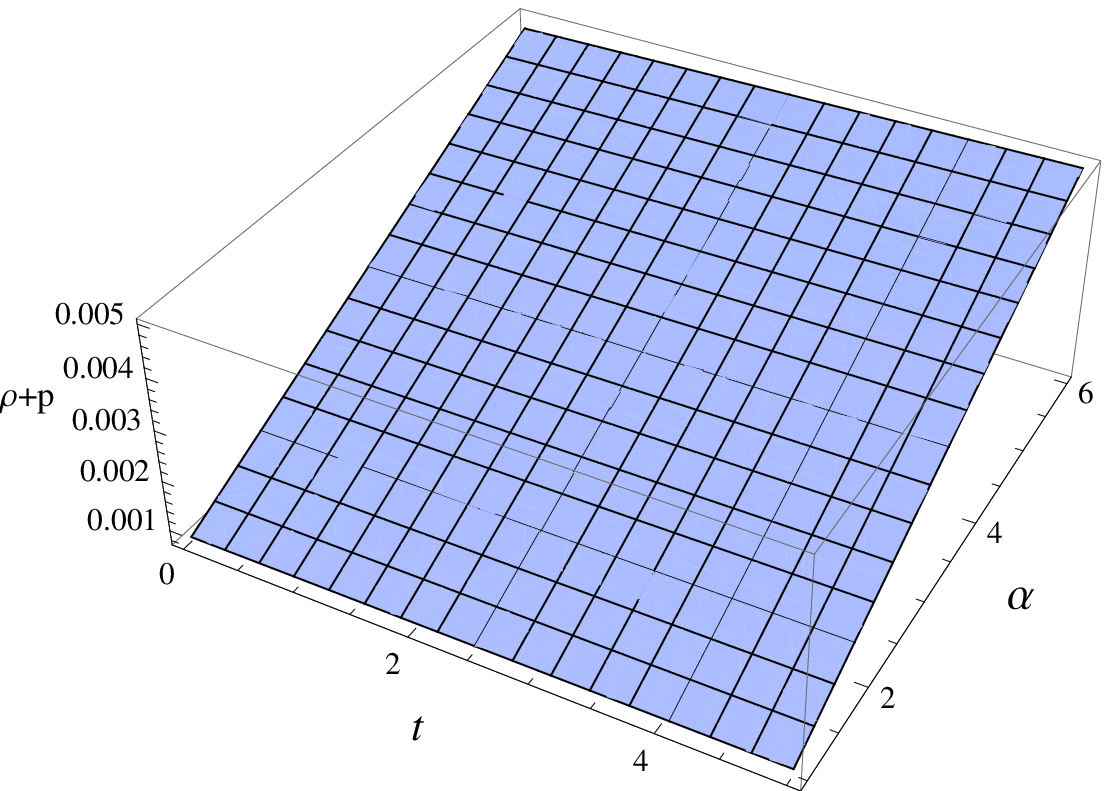,width=.50\linewidth,
height=2.2in} \caption{This figure is plotted for $n=2$. The left
part shows that NEC is violated for $\alpha<0$, whereas NEC is
satisfied for $\alpha>0$ shown on right side. We set
$l=\lambda=0.1$, $k=c_2=3$ and $m=0.9$. (Colour online)}
\end{figure}
\begin{figure}
\centering \epsfig{file=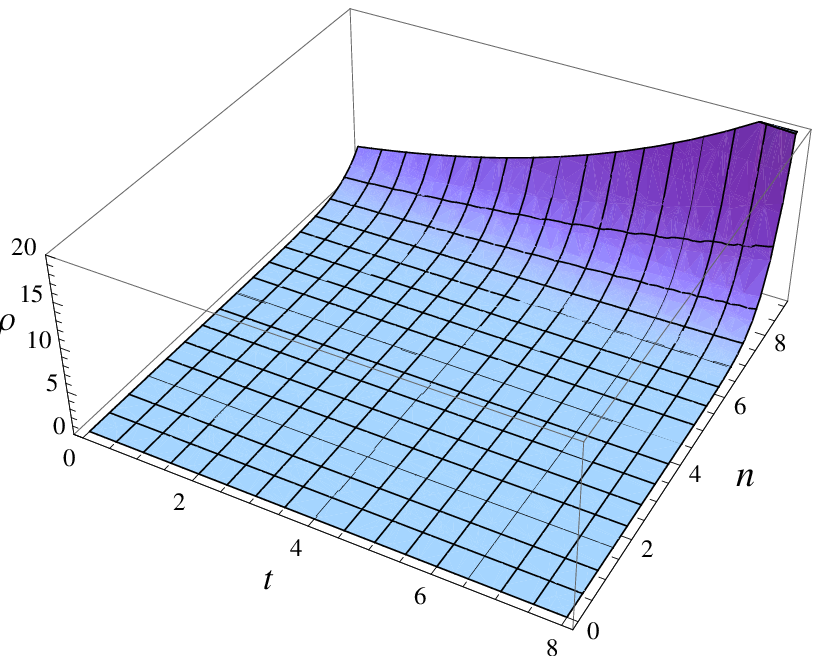} \caption{Plot of $\rho$ versus
cosmic time $t$ for $n\geq0$. We set $l=\lambda=0.1$, $k=c_2=3$,
$m=0.9$ and $\alpha=0.05$. (Colour online)}
\end{figure}

Equation (\ref{50}) shows that NEC is violated for the power law
expansion model. The behavior of NEC is shown in Figures
\textbf{5-6} which depends on the choice of $\alpha$ as well as $n$.
For each value of $n$ except $2\leq{n}\leq0$, NEC can be satisfied
for $\alpha<0$ but the choice $\alpha>0$ does not support it. If
$2\leq{n}\leq0$, the constraints to satisfy and violate NEC are
interchanged (see Figure \textbf{6}). For $m<1$, energy density
decreases in the range of $-6<n\leq0$ and increases with cosmic time
$t$ for $n>-7$. For $n>0$, the behavior of $\rho$ is shown in Figure
\textbf{7}. From Eqs.(\ref{5a}) and (\ref{15}), the Ricci scalar $R$
and $f(R,T)$ are given by
\begin{eqnarray*}
R&=&\frac{2}{3\alpha^2}\left[9l^2\alpha^2(m-2)(mlt+c_2)^{-2}-k^2(mlt+c_2)^
{-2(\frac{n+3}{m})}\right],\\\nonumber
f(R,T)&=&\frac{\alpha}{2}\left[R(mlt+c_2)^{\frac{n}{m}}+3n(n-m)l^2(mlt+c_2)^{\frac{n}{m}-2}\
\right.\\\nonumber&+&\left.9l(mlt+c_2)^{\frac{n}{m}-1}\right]+
\frac{8\pi(1-3\omega)+\lambda(1-\omega)}{6\alpha(8\pi+\lambda)(1+\omega)}
(2k^2(mlt+c_2)^{-\frac{n+6}{m}}\\\nonumber&+&3b_2(mlt+c_2)^{\frac{n}{m}-2}).
\end{eqnarray*}

\section{Massless Scalar Field Models}

The Lagrangian for massless scalar field $\phi$ is given
by$^{{20})}$
\begin{equation}\label{55}
\mathcal{L}_m=-\frac{1}{2}g^{\mu\nu}\partial_{\mu}\phi\partial_{\nu}\phi,
\end{equation}
and the corresponding energy-momentum tensor is
\begin{equation}\label{56}
T_{{\mu}{\nu}}=\partial_{\mu}\phi\partial_{\nu}\phi-\frac{1}{2}g_{\mu\nu}
\partial_{\gamma}\phi\partial^{\gamma}\phi.
\end{equation}
Here, $T_{{\mu}{\nu}}$ represents stiff matter with EoS
$\omega_{\phi}=1$. Using Eqs.(\ref{6}), (\ref{9}) and (\ref{56}), we
obtain the following field equations for massless scalar field
\begin{eqnarray}\label{57}
\left(\frac{\ddot{A}}{A}+2\frac{\ddot{B}}{B}-2\frac{\dot{A}\dot{B}}{AB}
-2\frac{\dot{B}^2}{B^2}\right)F+\frac{3}{2}\ddot{F}-\frac{1}{2}\left(\frac{\dot{A}}{A}
+2\frac{\dot{B}}{B}\right)\dot{F}=-\frac{3}{2}(8\pi+\lambda)\dot{\phi}^2,\\\label{58}
\left(\frac{\ddot{A}}{A}-2\frac{\ddot{B}}{B}+2\frac{\dot{A}\dot{B}}{AB}
-2\frac{\dot{B}^2}{B^2}\right)F-\frac{1}{2}\ddot{F}+\left(\frac{3}{2}\frac{\dot{A}}{A}
-\frac{\dot{B}}{B}\right)\dot{F}=\frac{1}{2}(8\pi+\lambda)\dot{\phi}^2,\\\label{59}
\left(\frac{\dot{B}^2}{B^2}-\frac{\ddot{A}}{A}\right)F-\frac{1}{2}\ddot{F}-
\frac{1}{2}\left(\frac{\dot{A}}{A}-2\frac{\dot{B}}{B}\right)\dot{F}
=\frac{1}{2}(8\pi+\lambda)\dot{\phi}^2.
\end{eqnarray}
\begin{figure}
\centering \epsfig{file=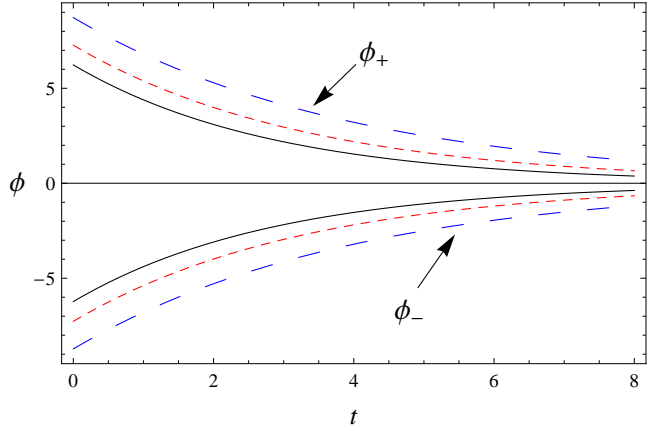} \caption{Evolution of $\phi$
versus cosmic time $t$ for $m=0$ and different values of $n$:
solid(black) $n=1$; dashed(red), $n=0$; dahsed(blue), $n=-1$. We set
$l=\lambda=0.1$, $k=3$ and $\alpha=0.05$. (Colour online)}
\end{figure}

\subsection{Scalar Field Models When $m=0$ and $m\neq0$}

The field equations with massless scalar field are similar to the
perfect fluid case with $\rho_{\phi}+p_{\phi}=\dot{\phi}^2$. Hence,
we obtain the same results for the scale factors and other physical
parameters. Substituting Eq.(\ref{30}) in (\ref{57})-(\ref{59}), we
obtain the time derivative of scalar field $\phi$ as
\begin{equation}\label{60}
\dot{\phi}=\pm\sqrt{\frac{-1}{3\alpha(8\pi+\lambda)}\left[2k^2e^{nlt-2(n+3)lt}+3b_1e^{nlt}\right]}.
\end{equation}
Using Eqs.(\ref{30}) and (\ref{60}) in (\ref{5a}), it follows that
\begin{equation}\label{61}
f(R,T)=\frac{\alpha}{2}(R+3l(n^2l+3))e^{nlt}-
\frac{4\pi}{3\alpha(8\pi+\lambda)}(2k^2e^{-(n+6)lt}+3b_1e^{nlt}).
\end{equation}
The behavior of $\phi$ for exponential expansion is shown in Figure
\textbf{8}. If we solve Eqs.(\ref{57})-(\ref{59}) for the case
$m\neq0$, we get the similar solutions as given in section
\textbf{2.2}. The expression of $\dot{\phi}$ is obtained as follows
\begin{equation}\label{62}
\dot{\phi}=\pm\sqrt{\frac{-1}{3\alpha(8\pi+\lambda)}\left[2k^2(mlt+c_2)^{-\frac{(n+6)}{m}}
+3b_2(mlt+c_2)^{\frac{(n-2m)}{m}}\right]}.
\end{equation}
Evolution of $\phi$ versus cosmic time $t$ for different values of
$n$ is shown in Figure \textbf{9}. Substituting Eqs.(\ref{43}) and
(\ref{62}) in Eq.(\ref{5a}), we have
\begin{eqnarray}\label{63}
&&f(R,T)=\frac{\alpha}{2}\left[R(mlt+c_2)^{\frac{n}{m}}+3n(n-m)l^2(mlt+c_2)^{\frac{n}{m}-2}\
+9l\right.\\\nonumber&\times&\left.(mlt+c_2)^{\frac{n}{m}-1}\right]-
\frac{4\pi}{3\alpha(8\pi+\lambda)}
\left[2k^2(mlt+c_2)^{-\frac{n+6}{m}}+3b_2(mlt+c_2)^{\frac{n}{m}-2}\right].
\end{eqnarray}
\begin{figure}
\centering \epsfig{file=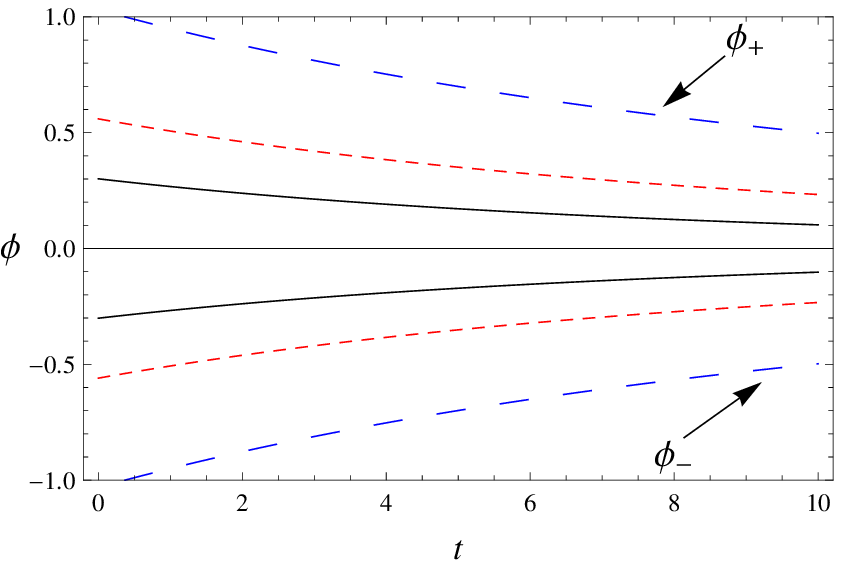} \caption{Evolution of $\phi$
versus cosmic time $t$ for $m\neq0$ and different values of $n$:
solid(black) $n=1$; dashed(red), $n=0$; dahsed(blue), $n=-1$. We set
$l=\lambda=0.1$, $k=c_2=3$, $m=0.9$ and $\alpha=0.05$. (Colour
online)}
\end{figure}

\section{Discussion and Conclusion}

The issue of accelerated expansion of the universe can be explained
by taking into account the modified theories of gravity such as
$f(R,T)$ gravity$^{{23})}$. In $f(R,T)$ gravity, cosmic acceleration
may result not only due to geometric contribution to the matter but
it also depends on matter contents of the universe. The coupling
between matter and geometry in this gravity results in nongeodesic
motion of test particles and an extra acceleration is always
present. This theory can be applied to explore several issues of
current interest and may lead to some significant results as
compared to other modified theories.

The homogeneous and anisotropic Bianchi universe models with perfect
fluid have been investigated in the context of $f(R,T)$
gravity$^{{14})}$. The exact solution of the field equations are
obtained for the particular choice $f(R,T)=R+2f(T)$ with
$f(T)={\lambda}T$. The results of this formulation are very similar
to that in general relativity and cannot imply the reconstruction of
$f(R,T)$ gravity$^{{14})}$. Houndjo$^{{15})}$ used a more general
form $f(R,T)=f(R)+{\lambda}T$ to reconstruct the $f(R,T)$ gravity
from holographic DE numerically. In present work, we have employed
this choice to reconstruct some explicit models of $f(R,T)$ gravity
for BI universe. We have presented the phantom evolution of the
universe by examining the NEC.

The exact solutions of the modified field equations are obtained for
the spatially homogeneous and anisotropic LRS BI universe with
perfect fluid and massless scalar field. The law of variation of
mean Hubble parameter is assumed which implies two cosmological
models for $m=0$ and $m\neq0$. These models represent the
accelerated expansion of the universe which supports the
observations of WMAP data and SNeIa$^{{1-2})}$. We have presented
physical properties of the models as well as kinematical parameters.
In the following, we summarize the results for these two models.
{\begin{itemize}
\item {Model for $V=c_1e^{3lt}$}
\end{itemize}}
For exponential expansion model, the accelerated expansion of the
universe may occur since $q=-1$. The kinematical paramters have been
discussed for two cases $n>-3$ and $n<-3$. The expansion scalar is
constant, while the Ricci scalar approaches to constant value as
$t\rightarrow{\infty}$ for $n>-3$ and take infinitely large values
for $n<-3$. The anisotropy parameter of expansion depends on time
and vanishes in future evolution for $n>-3$. If $\alpha>0$, NEC is
violated. In our discussion, we have considered a particular model
of $f(R,T)$ gravity, $f(R,T)=f(R)+\lambda{T}$. We are not able to
find the explicit function of $f(R,T)$ by using Eq.(\ref{5a}). For
$\lambda=0$, we develop $f(R)$ in terms of $R$ and hence the
function $f(R,T)$ as
\begin{eqnarray*}
f(R)=\left[\frac{2{\alpha}\omega}{1+\omega}(R+12H^2)+\frac{1}{2\alpha}
\{3{\alpha}^2H((n^2-4)H+3)\right.+\left.\frac{1-3\omega}
{1+\omega}b_1\}\right]e^{nlt},
\end{eqnarray*}
which can be expressed as
\begin{eqnarray*}
&&f(R_1)=const_1\times{R_1}^{m_1}+const_2\times{R_1}^{m_2},
\end{eqnarray*}
where $R_1=R+12H^2,~m_1=\frac{n+6}{2(n+3)}$ and $m_2=m_1-1$. The
models of $f(R_1)$ depending on $n$ are shown in table
\textbf{1}. We find that:\\\\
\textbf{Table 1:} Models of $f(R_1)$ corresponding to $n$.\\
\begin{table}[bht]
\centering
\begin{small}
\begin{tabular}{|c| c c c| }
\hline $n$  &&&$f(R_1)$\\
\hline $n=0$ &&&$R_1+R^0_1$, $R_1^{0}=const$\\
\hline$n=-2$ &&&$R_1+R_1^2$\\
\hline$n=-6$ &&&$R_1^0+\frac{1}{R_1}$\\
\hline$n=-4$ &&&$R_1^{-1}+R_1^{-2}$\\
\hline$n=-\frac{12}{5}$ &&&$R_1^3+R_1^2$\\
\hline$n=-\frac{18}{5}$ &&&$R_1^{-2}+R_1^{-3}$\\
\hline$n=-\frac{3}{2}$ &&&$R_1^\frac{3}{2}+R_1^{\frac{1}{2}}$\\
\hline$n=-\frac{9}{2}$ &&&$R_1^\frac{-1}{2}+R_1^\frac{-3}{2}$\\
\hline
\end{tabular}\end{small}
\end{table}\\
For $n=0,~f(R)$ represents the \emph{$\Lambda$CDM} model,
\emph{i.e.}, $f(R)=R+\Lambda$. If we put $constant=0$, then $f(R,T)$
is of the form $f(R,T)=R+T$. The most famous \emph{Starobinsky's}
model$^{{24})}$, $f(R)=R+\alpha{R}^2$ is achieved for $n=-2$ and the
corresponding $f(R,T)$ function is $f(R,T)=R+\alpha{R}^2+T$. For
$n=-6$, $f(R,T)$ can be presented as $f(R,T)=\frac{1}{R}+T$.

For massless scalar field $(m=0)$, we have found similar results for
scale factors as in perfect fluid. The expression of $f(R)$ is
\begin{eqnarray*}
&&f(R)=\left[{\alpha}(R+12H^2)+\frac{1}{2\alpha}
\{3{\alpha}^2H((n^2-4)H+3)-b_1\}\right]e^{nlt}.
\end{eqnarray*}
\emph{i.e.},
$f(R_1)=const_3\times{R_1}^{m_1}+const_4.\times{R_1}^{m_2}$.
{\begin{itemize}
\item {Model for $V=(mlt+c_2)^{3/m}$}
\end{itemize}}
For $m\neq0$, the deceleration parameter is $q=m-1$, which leads to
the accelerating universe model for $0<m<1$ and if $m>1(q>0)$, the
model represents decelerating phase of the universe. The evolution
of the scale factors is discussed for two cases $m>n+3$ and $m<n+3$
with $0<m<1$. The anisotropy parameter of expansion increases for
$n<-3$, whereas it may result in isotropic expansion in future
evolution of the universe for $n>-3$. The Hubble parameter,
expansion scalar and shear scalar approach to constant at earlier
times of the universe and approach to zero as
$t\rightarrow{\infty}$. The scalar curvature $R$ becomes constant as
$t\rightarrow{\infty}$ for $n>-3$, whereas it diverges for $n<-3$.
For $\lambda=0$, we have
\begin{eqnarray*}
f(R)&=&(mlt+c_2)^{n/m}\left[\frac{2{\alpha}\omega}{1+\omega}(R-6(m-2)H^2)
+\frac{\alpha}{2}H\{3H(n^2\right.\\\nonumber
&+&\left.m-4)+\frac{1-3\omega}
{\alpha^2l^2(1+\omega)}b_2H+9\}\right]
\end{eqnarray*}
which results in
\begin{eqnarray*}
&&f(R_2)=const_5\times{R_2}^{m_1}+const_6\times{R_2}^{m_2},
\end{eqnarray*}
where $R_2=R-6(m-2)H^2$. In case of massless scalar field, $f(R)$ is
\begin{eqnarray*}\nonumber
f(R)&=&(mlt+c_2)^{n/m}\left[\alpha(R-6(m-2)H^2)
+\frac{\alpha}{2}H\{3H(n^2+m\right.\\\nonumber
&-&\left.4)+\frac{b_2H} {\alpha^2l^2}+9\}\right],
\end{eqnarray*}
\emph{i.e.},
$f(R_2)=const_7\times{R_2}^{m_1}+const_8\times{R_2}^{m_2}$. The
scalar field $\phi$ is found to be decreasing function of cosmic
time (see Figures \textbf{8, 9})$^{{19})}$.

We have seen that all $f(R)$ represent identical behavior with
different constraints. The NEC is found to be violated for both
models $m=0$ and $m\neq0$ which results in phantom evolution. For
${\omega}<-1$, energy density is found to be positive and pressure
is negative. Thus, our solutions for perfect fluid represent the
phantom era of DE which are consistent with present observations of
WMAP5$^{{25})}$. The isotropic behavior of models is observed for
future evolution. It can be concluded that solutions of massless
scalar field in $f(R,T)$ gravity can be recovered for both
exponential and power law expansion models from phantom solutions of
perfect fluid if $\omega=1$.

\newpage

{\bf Acknowledgment}

\vspace{0.25cm}

We would like to thank the anonymous referee for fruitful
comments.\\\\
1) C. L. Bennett, M. Halpern, G. Hinshaw, N. Jarosik, A. Kogut, M.
Limon, S. S. Meyer, L. Page, D. N. Spergel, G. S. Tucker, E.
Wollack, E. L. Wright, C. Barnes, M. R. Greason, R. S. Hill, E.
Komatsu, M. R. Nolta, N. Odegard, H. V. Peiris, L. Verde and J. L.
Weiland: Astrophys. J. Suppl. \textbf{148} (2003) 1; D. N. Spergel,
L. Verde, H. V. Peiris, E. Komatsu, M. R. Nolta, C. L. Bennett, M.
Halpern, G. Hinshaw, N. Jarosik, A. Kogut, M. Limon, S. S. Meyer, L.
Page, G. S. Tucker, J. L. Weiland, E. Wollack and E. L. Wright:
Astrophys. J. Suppl. \textbf{148} (2003) 175; D. N. Spergel, R.
Bean, O. Doré, M. R. Nolta, C. L. Bennett, J. Dunkley, G. Hinshaw,
N. Jarosik, E. Komatsu, L. Page, H. V. Peiris, L. Verde, M. Halpern,
R. S. Hill, A. Kogut, M. Limon, S. S. Meyer, N. Odegard, G. S.
Tucker, J. L. Weiland, E. Wollack and E. L. Wright:
Astrophys. J. Suppl. \textbf{170} (2007) 377.\\\
2) S. Perlmutter, S. Gabi, G. Goldhaber, A. Goobar, D. E. Groom, I.
M. Hook, A. G. Kim, M. Y. Kim, G. C. Lee, R. Pain, C. R.
Pennypacker, I. A. Small, R. S. Ellis, R. G. McMahon, B. J. Boyle,
P. S. Bunclark, D. Carter, M. J. Irwin, K. Glazebrook, H. J. M.
Newberg, A. V. Filippenko, T. Matheson, M. Dopita and W. C. Couch:
Astrophys. J. \textbf{483} (1997) 565; S. Perlmutter, G. Aldering,
M. D. Valle, S. Deustua, R. S. Ellis, S. Fabbro, A. Fruchter, G.
Goldhaber, A. Goobar, D. E. Groom, I. M. Hook, A. G. Kim, M. Y. Kim,
R. A. Knop, C. Lidman, R. G. McMahon, P. Nugent, R. Pain, N.
Panagia, C. R. Pennypacker, P. Ruiz-Lapuente, B. Schaefer and N.
Walton: Nature \textbf{391} (1998) 51; S. Perlmutter, G. Aldering,
G. Goldhaber, R. A. Knop, P. Nugent, P. G. Castro, S. Deustua, S.
Fabbro, A. Goobar, D. E. Groom, I. M. Hook, A. G. Kim, M. Y. Kim, J.
C. Lee, N. J. Nunes, R. Pain, C. R. Pennypacker, R. Quimbey, C.
Lidman, R. S. Ellis, M. Irwin, R. G. Mcmahon, P. Ruiz-lapuente, N.
Walton, B. Schaefer, B. J. Boyle, A. V. Filippenko, T. Matheson, A.
S. Fruchter, N. Panagia, H. J. M. Newberg and W. J. Couch:
Astrophys. J. \textbf{517} (1999) 565; A. G. Riess, L. G. Strolger,
J. Tonry, Z. Tsvetanov, S. Casertano, H. C. Ferguson, B. Mobasher,
P. Challis, N. Panagia, A. V. Filippenko, W. Li, R. Chornock, R. P.
Kirshner, B. Leibundgut, M. Dickinson, A. Koekemoer, N. A. Grogin
and M. Giavalisco : Astrophys. J. \textbf{607} (2004) 665; A. G.
Riess, L. G. Strolger, S. Casertano, H. C. Ferguson, B. Mobasher, B.
Gold, P. J. Challis, A. V. Filippenko, S. Jha, W. Li, J. Tonry, R.
Foley, R. P. Kirshner, M. Dickinson, E. MacDonald, D. Eisenstein, M.
Livio, J. Younger, C. Xu, T. Dahlén and D. Stern : Astrophys. J.
\textbf{659}
(2007) 98.\\\
3) E. Hawkins, S. Maddox, S. Cole, O. Lahav, D. S. Madgwick, P.
Norberg, J. A. Peacock, I. K. Baldry, C. M. Baugh, J.
Bland-Hawthorn, T. Bridges, R. Cannon, M. Colless, C. Collins, W.
Couch, G. Dalton, R. D. Propris, S. P. Driver, S.P., G. Efstathiou,
R. S. Ellis, C.S. Frenk, K. Glazebrook, C. Jackson, B. Jones, I.
Lewis, S. Lumsden, W. Percival, B. A. Peterson, W. Sutherland and K.
Taylor: Mon. Not. Roy. Astron. Soc. \textbf{346} (2003) 78; M.
Tegmark, M. A. Strauss, M. R. Blanton, K. Abazajian, S. Dodelson, H.
Sandvik, X. Wang, D. H. Weinberg, I. Zehavi, N. A. Bahcall, F.
Hoyle, D. Schlegel, R. Scoccimarro, M. S. Vogeley, A. Berlind, T.
Budavari, A. Connolly, D. J. Eisenstein, D. Finkbeiner, J. A.
Frieman, J. E. Gunn, L. Hui, B. Jain, D. Johnston, S. Kent, H. Lin,
R. Nakajima, R. C. Nichol, J. P. Ostriker, A. Pope, R. Scranton, U.
Seljak, R. K. Sheth, A. Stebbins, A. S. Szalay, I. Szapudi, Y. Xu,
J. Annis, J. Brinkmann, S. Burles, F. J. Castander, I. Csabai, J.
Loveday, M. Doi, M. Fukugita, B. Gillespie, G. Hennessy, D. W. Hogg,
Z. E. Ivezic´, G. R. Knapp, D. Q. Lamb, B. C. Lee, R. H. Lupton, T.
A. McKay, P. Kunszt, J. A. Munn, L. Connell, J. Peoples, J. R. Pier,
M. Richmond, C. Rockosi, D. P. Schneider, C. Stoughton, D. L.
Tucker, D. E. V. Berk, B. Yanny and D. G. York: Phys. Rev. D
\textbf{69} (2004) 103501; S. Cole , W. J. Percival, J. A. Peacock,
P. Norberg, C. M. Baugh, C. S. Frenk, I. Baldry, J. B. Hawthorn, T.
Bridges, R. Cannon, M. Colless, C. Collins, W. Couch, N. J. G.
Cross, G. Dalton, V. R. Eke, R. D. Propris, S. P. Driver, G.
Efstathiou, R. S. Ellis, K. Glazebrook, C. Jackson, A. Jenkins, O.
Lahav, I. Lewis, S. Lumsden, S. Maddox, D. Madgwick, B. A. Peterson,
W. Sutherland and K. Taylor: Mon. Not. Roy.
Astron. Soc. \textbf{362} (2005) 505.\\\
4) D. J. Eisentein,  I. Zehavi, D. W. Hogg, R. Scoccimarro, M. R.
Blanton, R. C. Nichol, R. Scranton, Hee-Jong Seo, M. Tegmark, Z.
Zheng, S. F. Anderson, J. Annis, N. Bahcall, J. Brinkmann, S.
Burles, F. J. Castander, A. Connolly, I. Csabai, M. Doi, M.
Fukugita, J. A. Frieman, K. Glazebrook, J. E. Gunn, J. S. Hendry, G.
Hennessy, Z. Ivezic', S. Kent, G. R. Knapp, H. Lin, Yeong-Shang Loh,
R. H. Lupton, B. Margon, T. A. McKay, A. Meiksin, J. A. Munn, A.
Pope, M. W. Richmond, D. Schlegel, D. P. Schneider, K. Shimasaku, C.
Stoughton, M. A. Strauss, M. SubbaRao, A. S. Szalay, I. Szapudi, D.
L. Tucker, B. Yanny, and D. G. York: Astrophys. J. \textbf{633}
(2005) 560.\\\
5) B. Jain and A. Taylor: Phys. Rev. Lett. \textbf{91} (2003)
141302.\\\
6) P. Astier J. Guy, N. Regnault, R. Pain, E. Aubourg, D. Balam, S.
Basa, R. G. Carlberg, S. Fabbro, D. Fouchez, I. M. Hook, D. A.
Howell, H. Lafoux, J. D. Neill, N. Palanque-Delabrouille, K.
Perrett, C. J. Pritchet, J. Rich, M. Sullivan, R. Taillet, G.
Aldering, P. Antilogus, V. Arsenijevic, C. Balland, S. Baumont, J.
Bronder, H. Courtois, R. S. Ellis, M. Filiol, A. C. Gonçalves, A.
Goobar, D. Guide, D. Hardin, V. Lusset, C. Lidman, R. McMahon, M.
Mouchet, A. Mourao, S. Perlmutter, P. Ripoche, C. Tao and N. Walton
: Astron. Astrophys. \textbf{447} (2006) 31; S. Tsujikawa: Lect.
Notes Phys. \textbf{800} (2010) 99; P. J. E. Peebles: Rev. Mod.
Phys.
\textbf{75} (2003) 559.\\\
7) V. Sahni: Lect. Notes Phys. \textbf{653} (2004) 141; T.
Padmanabhan: Gen. Relativ. Gravit. \textbf{40} (2008) 529; R. R.
Caldwell: Phys. Lett. B \textbf{545} (2002) 23; S. Nojiri and S. D.
Odintsov: Phys. Lett. B \textbf{562} (2003) 147; B. Feng, X. L. Wamg
and X. M. Zhang: Phys. Lett. B \textbf{607} (2005) 35; A. A. Sen:
Phys. Rev. D \textbf{66} (2002)
043507; T. Padmanabhan: Phys. Rev. D \textbf{66} (2002) 021301.\\\
8) O. Akarsu and C. B. Kilinc: Gen. Relativ. Gravit. \textbf{42}
(2010) 1; M. Sharif and M. Zubair: Int. J. Mod. Phys. D \textbf{19}
(2010) 1957; Astrophys. Space Sci. \textbf{330}
(2010) 399; ibid. \textbf{339} (2012) 45.\\\
9) S. Nojiri and S. D. Odintsov: Int. J. Geom. Methods Mod. Phys.
\textbf{4} (2007) 115; T. P. Sotiriou and V. Faraoni: Rev.
Mod. Phys. \textbf{82} (2010) 451.\\\
10) R. Ferraro and F. Fiorini: Phys. Rev. D \textbf{75} (2007)
084031; G. R. Bengochea and R. Ferraro: Phys. Rev. D \textbf{79}
(2009) 124019; E. V. Linder: Phys. Rev. D \textbf{81} (2010)
127301.\\\
11) S. M. Carroll, A. De Felice, V. Duvvuri, D. A. Easson, M.
Trodden and M. S. Turner: Phys. Rev. D \textbf{71} (2005) 063513; G.
Cognola, E. Elizalde, S. Nojiri, S. D. Odintsov and S. Zerbini:
Phys. Rev. D \textbf{73} (2006) 084007.\\\
12) T. Harko, F. S. N. Lobo, S. Nojiri and S. D. Odintsov:
Phys. Rev. D \textbf{84} (2011) 024020.\\\
13) L. D. Landau and E. M. Lifshitz: \emph{The Classical Theory
of Fields} (Butterworth-Heinemann, 2002).\\\
14) K. S. Adhav: Astrophys. Space Sci. \textbf{339} (2012) 365; D.
R. K. Reddy, R. Santikumar and R. L. Naidu: Astrophys. Space Sci.
DOI: 10.1007/s10509-012-1158-7; R. Chaubey, A.K. Shukla: Astrophys.
Space Sci. DOI: 10.1007/s10509-012-1204-5.\\\
15) M. J. S. Houndjo: Int. J. Mod. Phys. D \textbf{21} (2012)
1250003; M. J. S. Houndjo and O. F. Piattella: Int. J. Mod. Phys. D
\textbf{21} (2012) 1250024; M. Jamil, D. Momeni, M. Raza and R.
Myrzakulov: Eur. Phys. J. C \textbf{72} (2012)
1999.\\\
16) M. Sharif and M. Zubair: JCAP \textbf{03} (2012) 028 [Errata
\textbf{05} (2012) E01]; \emph{Thermodynamic Behavior of $f(R,T)$
Gravity Models at the Apparent Horizon} (To appear in Cent. Eur. J.
Phys).\\\
17) M. Sharif and M. F. Shamir: Class. Quantum Grav. \textbf{26}
(2009) 235020; Gen. Relativ. Gravit. \textbf{42} (2010)
2643.\\\
18) M. Sharif and H. R. Kausar: Phys. Lett.
B \textbf{697} (2011) 1.\\\
19) M. Sharif and W. Saira: Eur. Phys. J. C \textbf{72} (2012) 1876;
C. Aktas, S. Aygun and I. Yilmaz: Phys.
Lett. B \textbf{707} (2012) 237.\\\
20) S. Capozziello: Phys. Lett. B \textbf{639} (2012) 135; S.
Capozziello and M. Laurentis: Phys. Rep. \textbf{509} (2011) 1; D.
Yunshuang, Z. Hongsheng and L. Xin-Zhou: Eur. Phys. J. C \textbf{71}
(2011) 1660; Y. Bisabr: Phys. Lett. B \textbf{690} (2011) 456; V.
Faraoni: Class. Quantum Grav. \textbf{26} (2009)
145014.\\\
21) M. Sharif and M. Zubair: Astrophys. Space Sci.
DOI: 10.1007/s10509-012-1169-4.\\\
22) M. S. Berman: Nuovo Cimento B \textbf{74} (1983) 182; M. S.
Berman and F. M. Gomide: Gen. Relativ.
Gravit. \textbf{20} (1988) 191.\\\
23) K. Bamba, S. Capozziello, S. Nojiri and S. D. Odintsov:
Astrophys. Space Sci. DOI: 10.1007/s10509-012-1181-8.\\\
24) A. A. Starobinsky: Phys. Lett. B \textbf{91} (1980)
99.\\\
25) E. Komatsu, J. Dunkley, R. Nolta, C. L. Bennett, B. Gold, G.
Hinshaw, N. Jarosik, D. Larson, M. Limon, L. Page, D. N. Sperge, M.
Halpern, R. S. Hill, A. Kogut, S. S. Meyer, G. S. Tucker, J. L.
Weiland, E. Wollack, and E. L. Wright:
Astrophys. J. Suppl. \textbf{180} (2009) 330.\\\
\end{document}